\documentclass[pra,aps,11pt]{revtex4}
\usepackage[dvips,bookmarks=false]{hyperref}
\usepackage{amsmath,mathrsfs}
\usepackage{graphicx,epsfig}
\usepackage{latexsym,amssymb}
\usepackage{epsf,longtable,array}
\usepackage{natbib,hyperref}

\begin{document}

\title{Dark Energy From Fifth Dimension}
\author{H. Alavirad \footnote{alavirad@gmail.com} and N. Riazi \footnote{riazi@physics.susc.ac.ir}}
\affiliation{Physics Department and Biruni Observatory, Shiraz
University, Shiraz 71454, Iran}
\begin{abstract}
Observational evidence for the existence of  dark energy is
strong. Here we suggest a model which is based on a modified
gravitational theory in 5D and interpret the  5th dimension as a
manifestation of dark energy in the  4D observable universe. We
also obtain an equation of state parameter which varies with time.
Finally, we match our model with observations by choosing the free
parameters of the model.
\par
\par
Keywords: dark energy, extra dimensions, Kaluza-Klein theory.
\end{abstract}

\maketitle

\section{Introduction}
Recent observations suggest that the Universe is not only
expanding, but also accelerating \cite{Perlmutter,Riess}. The
first candidate for dark energy was a cosmological constant. It
was originally introduced by Einstein in 1917 to achieve a static
universe. However, after Hubble observations which suggested that
the universe is not static, it was abandoned.
\par In particle physics,
the cosmological constant normally arises as an energy density of
the vacuum. The energy scale of the cosmological constant
($\Lambda$) should be much larger than that of the present Hubble
constant $H_0$, if it originates from the vacuum energy density.
This is the cosmological constant problem.
\par In addition to the cosmological constant proposal, there are a
lot of alternative routes which have been proposed to explain the
accelerated expansion of the Universe. Some of the most important
are as follows \cite{Copeland}:
 \par1-Quintessence
models\cite{Carroll1};
\par2-Scalar field
models\cite{Frieman};
\par3-Chameleon fields
\cite{Khoury};
\par4-K-essence  \cite{Armend};
\par5-Modified gravity
arising out of string theory  \cite{Dvali} or generalization of GR
\cite{Carroll2, Nojiri1};
\par6-Phantom dark energy \cite{cald}; \par....
\par The
Einstein field equations consist of two parts. The first part
contains the geometry and the second part contains the matter.
Some of dark energy (DE) models include the cosmological constant
and a scalar field which modify the rhs by introducing some extra
terms in it. There is another way: modifying the geometry (i.e.
the lhs of the Einstein's equations). The geometrical
modifications can arise from quantum effects such as higher
curvature corrections to the Einstein-Hilbert (EH) action. In
\cite{Bertotti} by introducing a quadratic term in R, an
inflationary solution in the early universe was obtained. However,
it was pointed out in \cite{Carroll2, Nojiri1} that the late time
acceleration can be realized by adding a term containing inverse
power of Ricci scalar to the EH action. The structure of this
paper is as follows: in section \ref{II}, we review the modified
gravity theories. In section \ref{III}, we discuss briefly the
Kaluza-Klein gravity and the induced matter in this theory. In
section \ref{IV}, we apply the STM formalism in CDTT model. Then
in section \ref{V}, we extract energy density and the pressure of
dark energy from STM formalism. Finally, in section \ref{VI}, we
will match our model with observations by choosing the free
parameters of the model.

\section{Modified Gravity\label{II}}
We first investigate a fairly general way  of modifying
gravitational theory which is known as f(R) gravity. The formalism
starts with the introduction of an action in the form:
\begin{equation}
S=\int\sqrt{-g}f(R)d^{4}x+\int\sqrt{-g}\mathcal{L}_{\mathcal{M}}d^{4}x\label{1}
\end{equation}
where f(R) is an arbitrary function of R such as  $R+\alpha
R^{2}$, $R+\alpha lnR$, $R+\frac{\alpha}{R}$ etc, and
$\mathcal{L}_{\mathcal{M}}$ is a matter Lagrangian density. By
varying the action (1) with respect to the metric $g_{\mu \nu}$,
one obtains the following field equations\cite{Nojiri1}:
\begin{equation}
\frac{1}{2}g_{\mu
\nu}f(R)-R_{\mu\nu}f^{\prime}(R)-\nabla_{\mu}\nabla_{\nu}f^{\prime}(R)-g_{\mu
\nu}\nabla^{2}f^{\prime}(R)+\frac{1}{2}T_{\mu \nu}=0 \label{2}
\end{equation}
where $\mu$ ,$\nu$=0...3 and a prime shows differentiate with
respect to $R$ . By taking the trace of (\ref{2}) in the absence
of matter and constant curvature  $(\triangledown_{\alpha}R=0)$
case, we find:
\begin{equation}
f(R)-\frac{1}{2}R f^{\prime}(R)=0 \label{2}
\end{equation}
In \cite{Carroll2} Carroll et.al. considered a f(R) function of
the form (CDTT model):
\begin{equation}
f(R)=R-\frac{\mu^4}{R}\label{3}
\end{equation}
where $\mu$ is a new parameter with dimension of $[time]^{-1}$. By
means of (\ref{2}), we obtain the following field equations for
CDTT model:
\begin{equation}
\frac{1}{2}g_{\mu \nu}(R-\frac{\mu^4}{R})-R_{\mu
\nu}(1+\frac{\mu^4}{R^2})-\mu^4(\nabla_{\mu}\nabla_{\nu}+g_{\mu
\nu}\nabla^{2})\frac{1}{R^2}+\frac{1}{2}T_{\mu \nu}=0\label{4}
\end{equation}

\par The constant-curvature vacuum solutions in 4D, for which
$\triangledown_{\mu}R=0$, satisfy $R=\pm\sqrt{3}\mu^{2}$. Thus one
finds the interesting result that the constant curvature vacuum
solutions in 4D are not Minkowski space but rather are de Sitter
or anti de Sitter spaces. Moreover in spherically symmetric case
(the black hole solutions), one obtains a Schwartzchild-de Sitter
black hole.
\par By considering a
perfect fluid case i.e:
\begin{equation}
T^{M}_{\mu\nu}=(\rho_{M}+p_{M})U_{\mu}U_{\nu}+p_{M}g_{\mu\nu}\label{6}
\end{equation}
where $ U^{\mu}$ is the fluid rest-frame four velocity, $\rho_{M}$
is the energy density, $p_{M}$ is the pressure and we consider an
equation of state of the form $p_{M}=\omega\rho_{M}$, also we take
the metric of the flat Robertson-Walker (in four dimension) form,
i.e. $ds^{2}=-dt^{2}+a(t)^{2}dx^{2}$. One obtains the following
equation:
\begin{equation}
3H^{2}-\frac{\mu^4}{12(\dot{H}+2H^2)^2}(2H\ddot{H}+15H^2\dot{H}+2{\dot{H}}^{2}+6H^4)=\rho_{M}\label{7}
\end{equation}
for the time-time component of the field equation and:
\begin{equation}
\dot{H}+\frac{3}{2}-\frac{\mu^4}{12(\dot{H}+2H^2)^2}(4\dot{H}+9H^2-a^2\partial_{0}\partial_{0}(\frac{1}{a^2})-
2a^2H\partial_{0}(\frac{1}{a^2}))=-\frac{1}{2}p_{M}\label{8}
\end{equation}
for the space-space component. Equation (\ref{7}) reduces to the
usual Friedman equation by setting $\mu=0$. In \cite{Carroll2}
Carroll et.al. obtained some solutions such as eternal de Sitter,
power law acceleration and future singularity in vacuum case. As
mentioned in the Introduction, the solutions of this theory
contain instabilities \cite{Dolgov1}.
\par In \cite{Dolgov} Dolgove and Kawasaki also find instability
in the matter section of the above modified gravity model. In
\cite{Faraoni} Faraoni suggested that generally a model is stable
if it satisfies the condition:$f^{\prime \prime}>0$ and is
unstable if $f^{\prime \prime}<0$.
\par Dicke in \cite{Dick} discussed the Newtonian limit in
singular nonlinear modified  gravity models $(i.e.  f(0)=\infty)$
and found that a model has Newtonian limit if it satisfies
$f^{\prime \prime}(R_0)=0$ where $R_0$ is the solution of
(\ref{2}). Therefore the f(R) model which is introduced in
(\ref{3}), hasn't Newtonian limit and is unstable.
\section{Kaluza-Klein gravity and induced matter}\label{III}
Kaluza unified electromagnetism with gravity  in $1921$ by
applying Einstein's general theory of relativity to  a five rather
than four-dimensional spacetime manifold \cite{overdiom}. The
Einstein equations, in five dimensions with no five dimensional
energy-momentum tensor, are:
\begin{equation}
\hat{G}_{A B}=0\label{13}
\end{equation}
or equivalently:
\begin{equation}
\hat{R}{_{A B}}=0\label{14}
\end{equation}
where hat denotes a five dimensional quantity and A,B=0\ldots4 and
$\hat{G}_{A B}\equiv \hat{R}_{A B}-\frac{1}{2}\hat{R}\hat{g}_{A
B}$ is the Einstein tensor, $\hat{R}_{A B}$ and
$\hat{R}=\hat{g}_{A B}\hat{R}^{A B}$ are the five dimensional
Ricci tensor and Ricci scalar and $\hat{g}_{A B}$ is the five
dimensional metric. The absence of matter in the five dimensional
universe is the Einstein's idea which says that the universe in
higher dimensions is empty.
\par The five dimensional Ricci tensor and Christoffel
symbols are defined in terms of metric exactly as in four
dimensions (minimal extension). Then everything depends on one's
choice for the form of the five dimensional metric. Kaluza
proposed the following metric:
\begin{equation}
\hat{g}{_{A B}}=
\begin{pmatrix}
g_{\alpha
\beta}+\kappa^{2}\phi^{2}A_{\alpha}A_{\beta}&\kappa\phi^{2}A_{\alpha}\\
\kappa\phi^{2}A_{\beta}&\phi^{2}\\
\end{pmatrix}\label{15}
\end{equation}
where the electromagnetic potential is scaled by a constant
$\kappa$ in order to get the right multiplicative factor in
action. \par There is an important question with the Kaluza's
assumption: where is the fifth dimension? Why do we not observe
it? Kaluza  proposed the cylindrical condition to overcome this
problem which means dropping all derivatives with respect to the
fifth coordinate.
\par In 1926 Klein showed that Kaluza's cylinder
condition would arise naturally if the fifth dimension has (1) a
circular topology, in which case physical fields would depend on
it only periodically, and could be Fourier-expanded; and (2) a
small enough ("compactified") scale in which case the energies of
all Fourier modes above the ground state could be made so high as
to be unobservable. This version of Klein theory is called
compactified Kaluza-Klein gravity.
\par If one applies the
cylinder condition and use the metric (\ref{15}) and field
equations (\ref{13}), then one will find that the $\alpha\beta-$,
$\alpha4-$ and $44$-components of the five dimensional field
equations (\ref{13}) reduce respectively to the following field
equations in four dimensions:
\begin{equation}
G_{\alpha \beta}=\frac{\kappa^{2}\phi^{2}}{2}T^{EM}_{\alpha
\beta}-\frac{1}{\phi}[\triangledown_{\alpha}(\partial_{\beta}\phi)-g_{\alpha
\beta}\Box{\phi}]\nonumber
\end{equation}
\begin{equation}
\triangledown^{\alpha}F_{\alpha
\beta}=-3\frac{\partial^{\alpha}\phi}{\phi}F_{\alpha
\beta}\nonumber
\end{equation}
\begin{equation}
\Box\phi=\frac{\kappa^{2}\phi^3}{4}F_{\alpha \beta}F^{\alpha
\beta} \label{16}
\end{equation}
where $G_{\alpha \beta}\equiv R_{\alpha \beta}-R g_{\alpha
\beta}/2$ is the Einstein tensor in four dimensions, $T^{EM}_{
\alpha \beta}\equiv g_{\alpha \beta}F_{\gamma \delta}\frac
{F^{\gamma \delta}}{4}-F^{\gamma}_{\alpha} F_{\beta \gamma}$ is
the electromagnetic energy-momentum tensor, and $F_{\alpha
\beta}\equiv\partial_{\alpha}A_{\beta}-\partial_{\beta}A_{\alpha}$,
 where $\alpha,\beta=0 \ldots 3$. If the scalar field $\phi$ is
constant throughout spacetime, then the first two equations of
(\ref{16}) are just the Einstein and Maxwell equations :
\begin{equation}
G_{\alpha \beta}=8\pi G\phi^{2}T^{EM}_{\alpha \beta}\nonumber
\end{equation}
\begin{equation}
\triangledown^{\alpha}F_{\alpha \beta}=0\label{17}
\end{equation}
where we have identified the scaling parameter $\kappa$ in terms
of the gravitational constant G (in four dimensions) by:
\begin{equation}
\kappa\equiv 4\sqrt{\pi G}\label{18}
\end{equation}
This is the result originally obtained by Kaluza and Klein, who
set $\phi=1$. The condition $\phi = constant.$ is consistent with
the third equation of (\ref{16}) when $ F_{\alpha \beta}F^{\alpha
\beta}=0,$ which was first pointed out by Jordan \cite{Jordan}.
\par An alternative is to abandon the cylindrical condition
\cite{Wesson1,Wesson2}. Therefore the metric depends on the fifth
dimension and this dependence allows one to obtain electromagnetic
radiation, dust and other forms of cosmological matter. These
types of theories are called noncompactified Kaluza-Klein
theories.
\par Wesson proposed that the fifth coordinate $\psi$
might be related to the rest mass. Dimensionally
$x^{4}=\frac{Gm}{c^2}$ allows us to treat the rest mass m of a
particle as a length coordinate, in analogy with $x^{0}=ct$. We
can say that the four-dimensional matter is a manifestation of the
five-dimensional geometry \cite{Wesson1}.
\par In noncompactified Kaluza-Klein theories we begin with a
metric of the form:
\begin{equation}
\hat{g}_{A B}=
\begin{pmatrix}
g_{\alpha \beta}&0\\
0&\epsilon\phi^2&\\
\end{pmatrix}
\end{equation}\label{19}
where we have introduced the factor $\epsilon$ in order to allow a
timelike or a  spacelike signature for the fifth dimension (and
$\epsilon^2=1$).
\par  Now by  using the five dimensional field equations
(\ref{13}) in vacuum and keeping derivatives with respect to the
fifth coordinate $x^4$, the resulting expression  for the $\alpha
\beta-$$\alpha4-$ and $44-$ parts of the five dimensional Ricci
tensor $R_{\alpha \beta}$ are :
\begin{eqnarray}
\hat{R}_{\alpha \beta}&=&R_{\alpha
\beta}-\frac{\triangledown_{\beta}(\partial_{\alpha}\phi)}{\phi}+
\frac{\epsilon}{2\phi^2}(\frac{\partial_{4}\phi
\partial_{4}g_{\alpha \beta}}{\phi}-\partial_{4}g_{\alpha \beta} \nonumber \\
&+& g^{\gamma \delta}\partial_{4}g_{\alpha
\gamma}\partial_{4}g_{\beta \delta}-\frac{g^{\gamma
\delta}\partial_{4}g_{\gamma \delta}\partial_{4}g_{\alpha
\beta}}{2})\label{20}
\end{eqnarray}
\begin{eqnarray}
\hat{R}_{\alpha 4}&=&\frac{g^{4 4}g^{\beta \gamma}}{4}
(\partial_{4}g_{\beta \gamma}\partial_{4}g_{4
4}-\partial_{\gamma}g_{4 4}\partial_{4}g_{\alpha
\beta})+\frac{\partial_{\beta}g^{\beta
\gamma}\partial_{4}g_{\gamma \alpha}}{2}  \nonumber \\
&&+\frac{g^{\beta \gamma}\partial_{4}(\partial_{\beta}g_{\gamma
\alpha})}{2}-\frac{\partial_{\alpha}g^{\beta
\gamma}\partial_{4}(g_{\beta \gamma})}{2}-\frac{g^{\beta
\gamma}\partial_{4}\partial_{\alpha}g_{\beta \gamma}}{2}
  \\ \nonumber   &&+\frac{g^{\beta \gamma}g^{\delta
\epsilon}\partial_{4}g_{\gamma \alpha}\partial_{\beta}g_{\delta
\beta}}{4}+\frac{\partial_{4}g^{\beta
\gamma}\partial_{\alpha}g_{\beta \gamma}}{4}\label{20.1}
\end{eqnarray}
\begin{eqnarray}
\hat{R}_{4 4}&=&-\epsilon\phi\Box\phi-\frac{\partial_{4}g^{\alpha
\beta}\partial_{4}g_{\alpha \beta}}{2}-\frac{g^{\alpha
\beta}\partial_{4}(\partial_{\alpha
\beta})}{2}+\frac{\partial_{4}\phi g^{\alpha
\beta}\partial_{4}g_{\alpha \beta}}{2\phi}\\ \nonumber
&-&\frac{g^{\alpha \beta}g^{\gamma \delta}\partial_{4}g_{\gamma
\beta}\partial_{4}g_{\alpha \delta}}{4}\label{20.2}
\end{eqnarray}
\par Then Eq. (\ref{20}) gives the following expression for the
four dimensional Ricci tensor:
\begin{eqnarray}
R_{\alpha
\beta}&=&\frac{\triangledown_{\beta}(\partial_{\alpha}\phi)}{\phi}-
\frac{\epsilon}{2\phi^2}[\frac{\partial_{4}\phi \partial_4
g_{\alpha \beta}}{\phi}-\partial_4 (\partial_4 g_{\alpha
\beta})\nonumber \\
&+& g^{\gamma \delta}\partial_4 g_{\alpha \gamma}\partial_4
g_{\beta \delta}-\frac{g^{\gamma \delta}\partial_{4}g_{\gamma
\delta}\partial_{4}g_{\alpha \beta}}{2}]\label{21}
\end{eqnarray}
\par The above equation allows us to interpret the four
dimensional matter as a manifestation of the five dimensional
geometry. We assume that the Einstein Field equations hold in four
dimensions i.e.:
\begin{equation}
8\pi GT_{\alpha \beta}=R_{\alpha \beta}-\frac{1}{2}Rg_{\alpha
\beta}\label{22}
\end{equation}
 where $T_{\alpha \beta}$ is the four-dimensional matter energy
 momentum tensor. By contracting (\ref{21}) with $g_{\alpha
 \beta}$,
 we obtain the following expression for Ricci scalar:
 \begin{equation}
 R=\frac{\epsilon}{4\phi^4}[\partial_{4}g^{\alpha \beta}\partial_{4}g_{\alpha \beta}
 +(g^{\alpha \beta}\partial_{4}g_{\alpha \beta})^{2}]\label{23}
 \end{equation}
 inserting (\ref{23}) and (\ref{21}) into (\ref{22}) one finds:
\begin{eqnarray}
8\pi G T_{\mu \nu}&=&\frac{\triangledown_\beta (\partial_\alpha
 \phi)}{\phi}-\frac{\epsilon}{2\phi^2}[\frac{\partial_4 \phi \partial_{4}g_{\alpha \beta}}{\phi
 }-\partial_{4}
  (\partial_{4}g_{\alpha \beta})+g^{\gamma \delta}\partial_{4}
 g_{\alpha \gamma}g_{\beta \gamma} \nonumber \\
 &-&\frac{g^{\alpha \gamma}\partial_{4}g_{\gamma \delta}\partial_4 g_{\alpha \beta}}{2}+\frac
 {g_{\alpha \beta}}{4}(\partial_4 g^{\gamma \delta}\partial_{4} g_{\gamma \delta}+(g^{\gamma \delta}\partial_4
 g_{\gamma \delta})^2)] \label{24}
\end{eqnarray}
If we use this expression for $T_{\mu \nu}$, the four-dimensional
Einstein equations $G_{\alpha \beta}=8\pi G T_{\alpha \beta}$ are
contained in the five-dimensional vacuum ones $\hat{G}_{A B}=0$.
The matter described by $T_{\mu \nu}$ is a manifestation of pure
geometry in the five dimensional world. There are solutions for
different types of metric and energy-momentum tensor, such as the
spherically symmetric case \cite{Wesson2}, the isotropic and
homogenous case \cite{Leon}, etc.
\section{5D modified gravity in STM Formalism }\label{IV}
It is easy to check that all f(R) gravity theories in the vacuum
and constant curvature case are equivalent to the Einstein field
equations in the presence of a cosmological constant, so we can
use any f(R) model. In the constant curvature and no matter case,
Eq.(\ref{2}) reduces to :
\begin{equation}
R_{\mu \nu}f^{\prime}(R)-\frac{1}{2}g_{\mu \nu}f(R)=0\label{24a}
\end{equation}
From (\ref{3}) we obtain $R_0$ and by substituting it in f(R) and
its derivative, and theb by replacing it in (\ref{24a}), the
following relation is obtained:
\begin{equation}
R_{\mu \nu}+\beta(R_0)g_{\mu \nu}=0\label{24b}
\end{equation}
where:
\begin{equation}
\beta(R_0)=-\frac{f(R_0)}{f^{\prime}(R_0)}\label{24c}
\end{equation}
 which can be rewritten as Einstein field equation in the presence
 of a cosmological constant, where:
 \begin{equation}
 \Lambda=\beta+\frac{R_0}{2}\label{24d}
 \end{equation}
In this section we use the CDTT model and work with a 5D extension
of the flat RW metric in the form:
\begin{equation}
ds^2=dt^2-a(t)^2({dr}^2+r^2{d\Omega}^2)-{R(t)}^2{d\psi}^2\label{25}
\end{equation}
where a(t) is the scale factor of ordinary 3D spatial dimensions
and R(t) is the scale factor of the fifth dimension. In five
dimensions, the solution of (\ref{2}) for CDTT model is (with
$M_{P}\equiv (8\pi G)^{-\frac{1}{2}}=1$):
\begin{equation}
\hat{R}=\pm \sqrt{\frac{7}{3}}\mu^2\label{26}
\end{equation}
Here we choose the minus sign.
\par Now we try to find out a(t) and R(t). By replacing (\ref{25}) and
(\ref{26}) in (\ref{24a}) in five dimensinal and vacuum (STM
formalism) and constant curvature case i.e:
\begin{equation}
H_{A B}= (1+\frac{{\mu}^{4}}{{\hat{R}}^2}){\hat{R}}_{A
B}-\frac{1}{2}(1-\frac{{\mu}^{4}}{{\hat{R}}^2}){\hat{R}}{\hat{g}}_{A
B}=0\label{27}
\end{equation}
 we obtain the following equations:
\begin{equation}
H^{0}_{0}=-15\frac{\ddot a(t)}{a(t)}-5\frac{\ddot
R(t)}{R(t)}+\frac{\sqrt{21}}{3}\mu^2=0\label{28}
\end{equation}
\begin{equation}
H^{1}_{1}=H^{2}_{2}=H^{3}_{3}= -5\frac{\ddot a(t)}{a(t)}-
10(\frac{\dot a(t)}{a(t)})^2-5\frac{\dot a(t)}{a(t)}\frac{\dot
R(t)}{R(t)}+\frac{\sqrt{21}}{3}\mu^2=0\label{29}
\end{equation}
\begin{equation}
H^{4}_{4}=-5\frac{\ddot R(t)}{R(t)}-15\frac{\dot
a(t)}{a(t)}\frac{\dot
R(t)}{R(t)}+\frac{\sqrt{21}}{3}\mu^2=0\label{30}
\end{equation}
By computing $\hat{R}$ for metric (\ref{25}) and replacing it in
(\ref{26}) we find:
\begin{equation}
6\frac{\ddot a(t)}{a(t)}+2\frac{\ddot R(t)}{R(t)}+6(\frac{\dot
a(t)}{a(t)})^2+6\frac{\dot a(t)}{a(t)}\frac{\dot
R(t)}{R(t)}-\frac{\sqrt{21\mu^2}}{3}=0\label{31}
\end{equation}
by solving simultaneously
(\ref{28}),(\ref{29}),(\ref{30}),(\ref{31}) we find a(t) and R(t)
:
\begin{equation}
a(t)=\pm \frac{\sqrt {-7\mu e^{\frac{\mu t
\sqrt{5}189^{\frac{1}{4}}}{15}}\sqrt{5}189^{\frac{3}{4}}(e^{\frac{{2\mu
t \sqrt{5}189^{\frac{1}{4}}}}{15}}C_2 -C_3)}}{\mu e^{\frac{\mu t
\sqrt{5} 189^{\frac{1}{4}}}{15}}}\label{32}
\end{equation}
\begin{equation}
R(t)=C_1\frac{(e^{\frac{4\mu t
\sqrt{5}3^{\frac{3}{4}}7^{\frac{1}{4}}}{15}}C_2^{2}-C_3^{2})^{\frac{1}{4}}(e^{\frac{2\mu
t
\sqrt{5}3^{\frac{3}{4}}7^{\frac{1}{4}}}{15}}C_2+C_3)^{\frac{3}{4}}}{(e^{\frac{2\mu
t
\sqrt{5}3^{\frac{3}{4}}7^{\frac{1}{4}}}{15}}C_2-C_3)^{\frac{3}{4}}(e^{\frac{4\mu
t
\sqrt{5}3^{\frac{3}{4}}7^{\frac{1}{4}}}{15}})^{\frac{1}{8}}}\label{33}
\end{equation}
In (\ref{32} ),(\ref{33}) we have three free parameters:
$C_1$,$C_2$,$C_3$. In  section \ref{IV} we will try to determine
them from the available observational parameters.
\section{Density and Pressure From STM }\label{V}
We consider a perfect fluid energy momentum tensor for dark energy
in four dimensions of the form:
\begin{equation}
T^{\mu}_{\nu}=diag(\rho_{DE}(t),-p_{DE}(t),-p_{DE}(t),-p_{DE}(t))
\end{equation}\label{34}
Wesson \cite{Wesson} suggested that the new terms due to fifth
dimension which depend on R(t) (and $\psi$ in noncompactified
Kaluza-Klein cosmology) in $H_{0}^{0}$ and $H_{a}^{a}$ (where a=1,
2,3), are density and pressure of matter respectively. The main
proposal of our model is to choose them as density and pressure of
the dark energy. By using equations (\ref{28}) and (\ref{29}), we
obtain the following expressions  for $\rho_{DE}(t)$ and
$p_{DE}(t)$:
\begin{equation}
\hat{H_0^0}=f(a(t))+\rho_{DE}(t)\label{35}
\end{equation}
\begin{equation}
\hat{H_a^a}=g(a(t))-p_{DE}(t)\label{36}
\end{equation}
and:
\begin{equation}
\rho_{DE}(t)=\frac{10}{7}\frac{\ddot R(t)}{R(t)}\label{37}
\end{equation}
\begin{equation}
p_{DE}(t)=-\frac{10}{7}\frac{\dot a(t)}{a(t)}\frac{\dot
R(t)}{R(t)}\label{38}
\end{equation}
It is clear that $\rho_{DE}(t)$ and $p_{DE}(t)$ vanish  for a
model with constant R(t).
\section{Typical values for the constants \label{VI}}
\par To specify $C_2$, $C_3$ and $\mu$ we appeal to the observational value of some cosmological parameters.
 From recent observations,
we know that the Hubble constant at present time is about $73\pm 8
 {km}s^{-1}{Mpc^{-1}}$ or in SI :
\begin{equation}
H(t_0)=\frac{\dot{a}(t)}{a(t)}|_{t_0}\backsimeq 23.6\pm3\times
10^{-19} s^{-1}\label{39}
\end{equation}
We also know that:
\begin{equation}
H(t)=\frac{\dot a(t)}{a(t)}\label{40}
\end{equation}\
The other cosmological parameter is the density parameter
$\Omega=\frac{8 \pi G \rho}{3H_{0}^{2}}$. Recent observations
suggest that the $\Omega_{tot}=\Omega_{DE}+\Omega_{m}\simeq 1$ and
the $\Omega_{DE}\simeq 0.7$. Another cosmological parameter is the
equation of state parameter (EoS) $w(t)=\frac{p(t)}{\rho(t)}$.
Recent observations limit the value of $\omega_{DE}$ between
$-0.4$ and $-1.02$. Another important cosmological parameter is
the transition redshift $z_T$, in which the cosmic evaluation
transfers from the decelerated era to the accelerated era or
equivalently the value of $w_{DE}$ is getting less than $-0.3$.
Observations suggest that the value of $z_T$ is between $0.15$ to
$0.5$ \cite{Gong}. We obtain these cosmological parameters for our
model and by this means we can get some idea about the free
parameters of the model. Here, we make five choices and discuss
them separately, in order to get an idea about the behavior of the
solutions. The constant $C_1$ can be determined by observations on
variation of some parameters during the history of the Universe
such as the fine structure constant $\alpha$ \cite{Stefanescu}.
\subsection{ Choice 1: $C_2=-1$,$C_3=0$, $\mu = 8.5\times 10^{-18}s^{-1}$}
In this case $w_{DE}(t)$ and $\Omega_{DE}(t)$ both are constant :
\begin{equation}
\ w_{ED}(t)=-1 \quad    \Omega_{DE}(t)= 0.47
\end{equation}
This case corresponds to a cosmological constant, but
$\Omega_{DE}(t)$ does not match with the observations. In this
case $H_{0}=0.23\times 10^{-17}s^{-1}$ in SI, which is in line
with observations.
\subsection{Choice 2: $C_2=-1$, $C_3=-1$, $\mu = 7\times 10^{-18}s^{-1} $}
\begin{figure}[h]
\begin{minipage}{8.0 cm}
\includegraphics[width=8.0 cm]{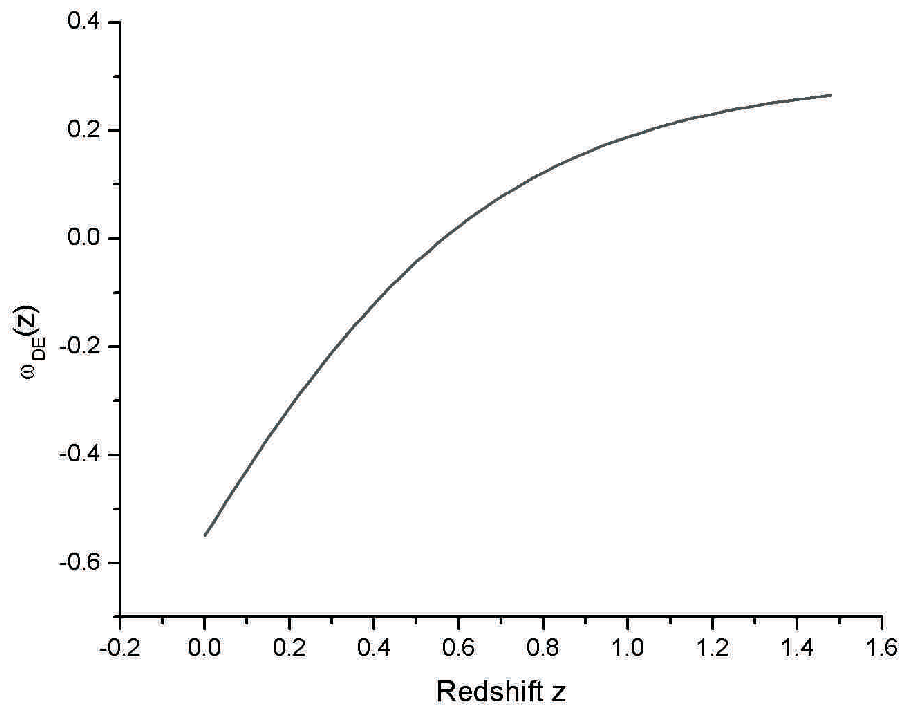}
\end{minipage}
\hfill
\begin{minipage}{8.0 cm}
\includegraphics[width=8.0 cm]{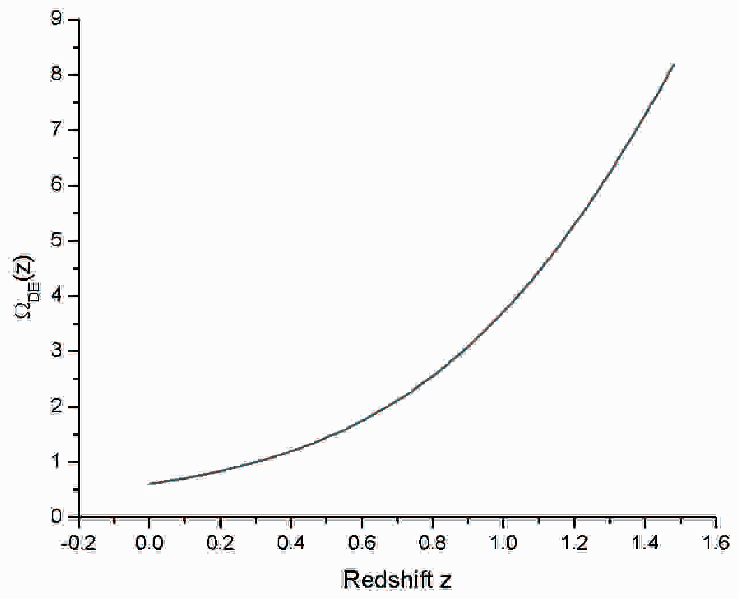}
\end{minipage}
\caption{$w_{DE}(t)$ (left) and $\Omega_{DE}(t)$ (right) for the
case $C_2=-1$ and $C_3=-1$,$\mu=7\times
10^{-18}s^{-1}$}\label{fig1}
\end{figure}

We plot $w_{DE}(t)$  and $\Omega_{DE}(t)$ for this choice in
Fig.\ref{fig1}. In this case at $z=0$ we have :
\begin{equation}
\omega_{ED}(0)=-0.55 \quad    \Omega_{DE}(0)= 0.61
\end{equation}
Observations suggest that at the present time ($z=0$);
$\Omega_{0}=\Omega_{DE0}+\Omega_{m0}\simeq 1$ so we obtain
$\Omega_{m0}= 0.39$.
In this case $H_{0}=0.23\times 10^{-17}s^{-1}$ in SI, which is in
line with observations. The value of $z_T$ in this case is $0.2$.
\subsection{Choice 3: $C_2=-9$, $C_3=-9$, $\mu=6\times 10^{-18}s^{-1}$}
\begin{figure}[h]
\begin{minipage}{8.0 cm}
\includegraphics[width=8.0 cm]{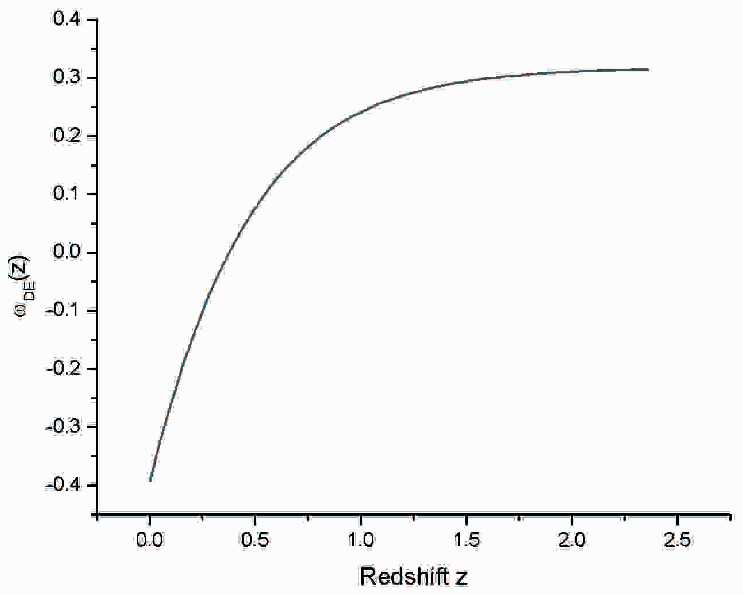}
\end{minipage}
\hfill
\begin{minipage}{8.0 cm}
\includegraphics[width=8.0 cm]{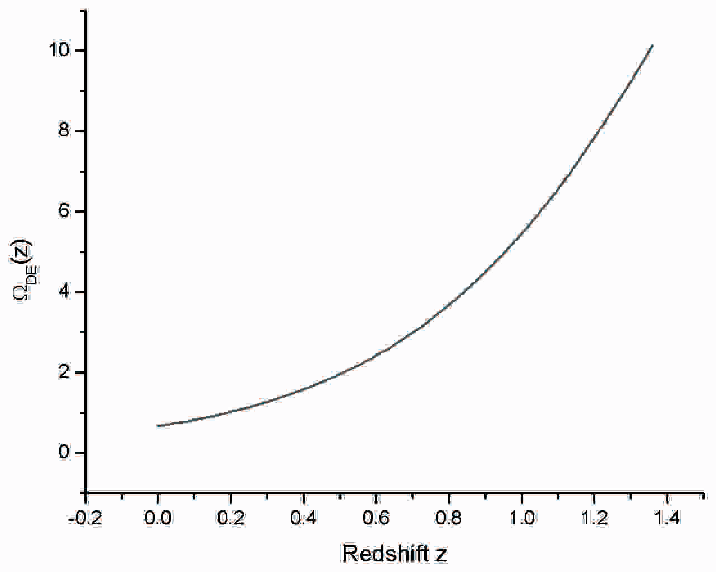}
\end{minipage}
\caption{$w_{DE}(t)$ (left) and $\Omega_{DE}(t)$ (right) for the
case $C_2=-9$ and $C_3=-9$,$\mu=6\times
10^{-18}s^{-1}$}\label{fig2}
\end{figure}

We plot the $w_{DE}(t)$ and $\Omega_{DE}(t)$ in Fig.\ref{fig2}. In
this case at $z=0$ we have :
\begin{equation}
\ w_{ED}(0)=-0.39 \quad    \Omega_{DE}(0)= 0.68
\end{equation}
In this case $H_{0}=0.19\times 10^{-17}s^{-1}$ in SI, which is in
agreement with observations. The value of $z_T$ in this case is
$0.1$.
\subsection{Choice 4: $C_2=-3$, $C_3=-6$, $\mu$ = $8\times 10^{-18}s^{-1}$}
\begin{figure}[h]
\begin{minipage}{8.0 cm}
\includegraphics[width=8.0 cm]{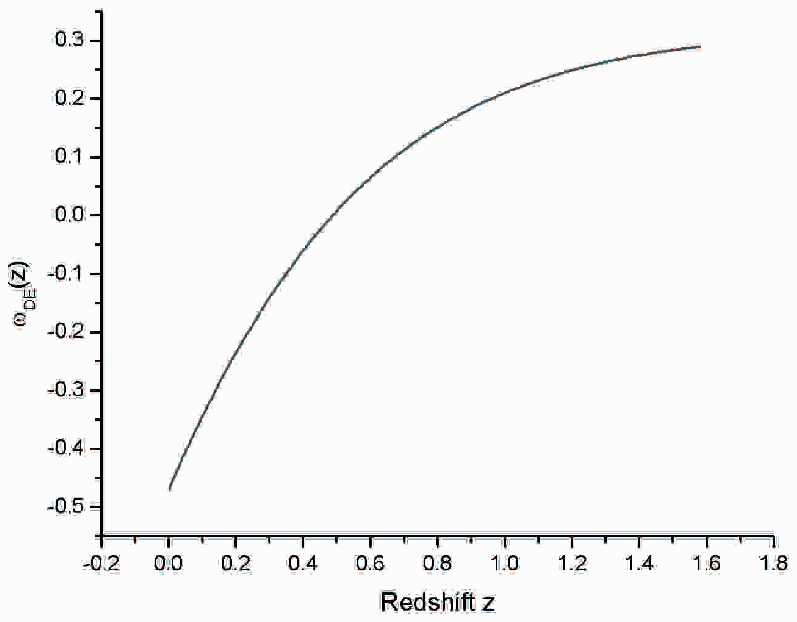}
\end{minipage}
\hfill
\begin{minipage}{8.0 cm}
\includegraphics[width=8.0 cm]{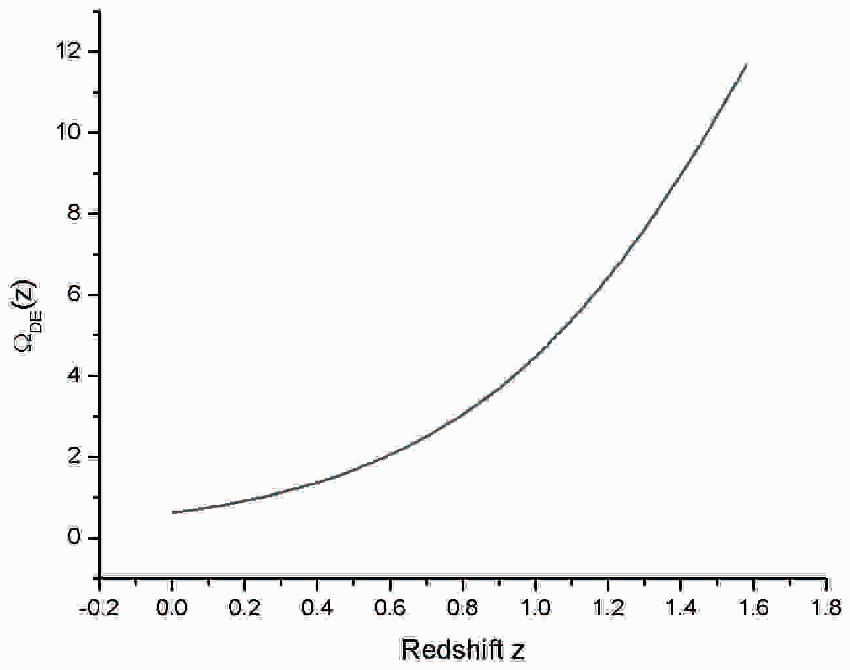}
\end{minipage}
\caption{$w_{DE}(t)$ (left) and $\Omega_{DE}(t)$ (right) for the
case $C_2=-3$ and $C_3=-6$,$\mu=8\times 10^{-18}s^{-1}$
}\label{fig3}
\end{figure}

We plot the $w_{DE}(t)$ and $\Omega_{DE}(t)$ in Fig.\ref{fig3}. In
this case at $z=0$ we have :
\begin{equation}
\ w_{ED}(0)=-0.47 \quad    \Omega_{DE}(0)= 0.64
\end{equation}

In this case $H_{0}=0.24\times 10^{-17}s^{-1}$ in SI, which is in
line with observations. The value of $z_T$ in this case is $0.15$.
\subsection{Choice 5: $C_2=-4$, $C_3=-9$, $\mu = 7\times 10^{-18}s^{-1}$}
\begin{figure}[h]
\begin{minipage}{8.0 cm}
\includegraphics[width=8.0 cm]{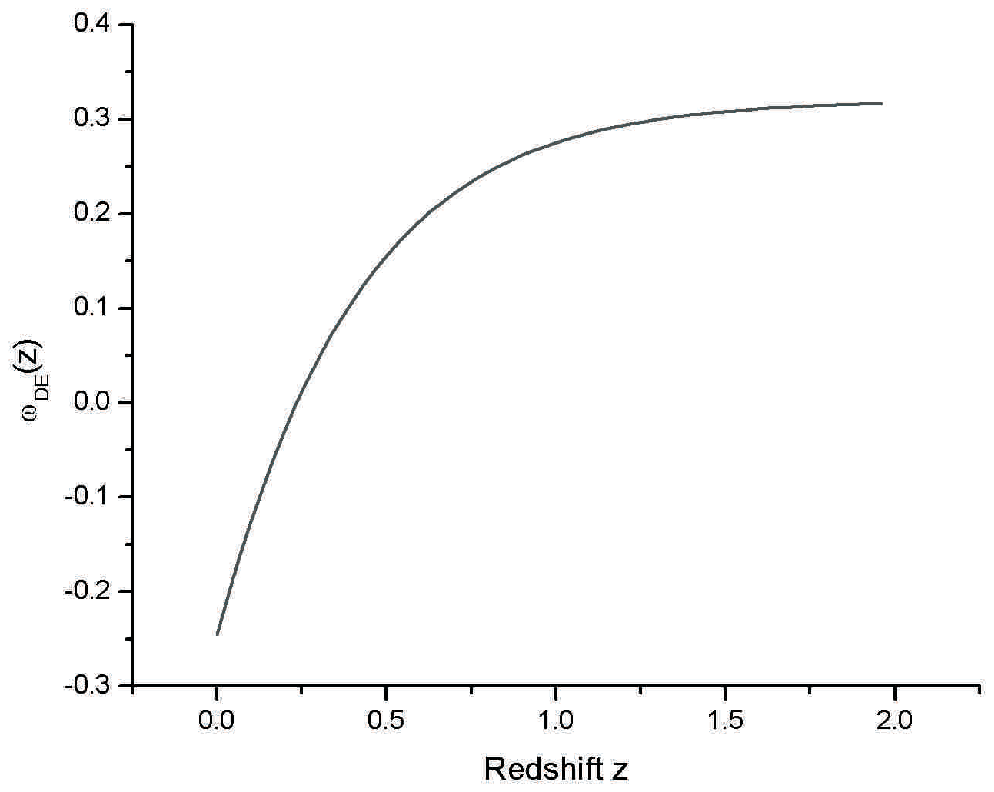}
\end{minipage}
\hfill
\begin{minipage}{8.0 cm}
\includegraphics[width=8.0 cm]{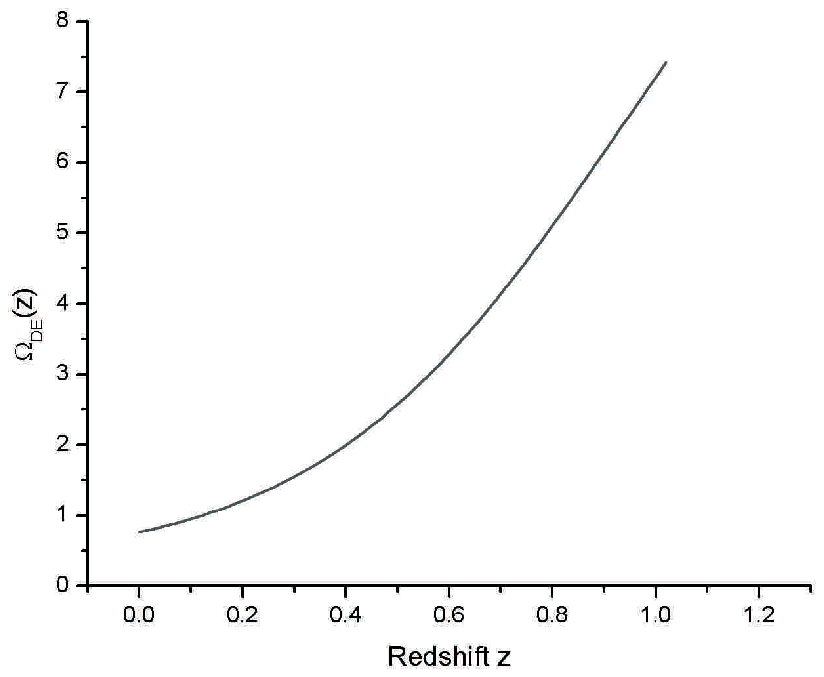}
\end{minipage}
\caption{$w_{DE}(t)$ (left) and $\Omega_{DE}(t)$ (right) for the
case $C_2=-4$ and $C_3=-9$,$\mu=7\times 10^{-18}s^{-1}$
}\label{fig4}
\end{figure}

We plot the $w_{DE}(t)$ and $\Omega_{DE}(t)$ in Fig.\ref{fig4} .
We have at $z=0$:
\begin{equation}
\ w_{ED}(0)=-0.25 \quad    \Omega_{DE}(0)= 0.75
\end{equation}

In this case $H_{0}=0.24\times 10^{-17}s^{-1}$ in SI, which is in
agreement with observations and the value of $w_{DE}$ at the
present time is more than $-0.3$, so we don't obtain an
accelerating universe.

\section{conclusion}
In this paper, we proposed a new model for dark energy by using
the CDTT f(R) gravity model and applying the STM formalism to a
five dimensional metric and interpret the fifth dimension as  dark
energy source. Then we used some cosmological parameters and
adjusted the model with observations to find the typical values
for the free parameters of the model ($C_2$, $C_3$ and $\mu$).
When we want to approach $\Omega_{DE}=0.7$ the equation of state
parameter tends to less than $-1/3$. In the forth case considered,
the cosmological parameters are close to the results of
observations, although this does not lead to a unique choice of
the parameters. If we take $C_3=0$, the model reduces to the
cosmological constant model.


\end{document}